\documentclass [pre, showpacs] {revtex4}
\usepackage{graphicx}

\begin{document}

\title{Phase transition in fiber bundle models with recursive dynamics}

\author{Pratip Bhattacharyya}
 \email{pratip@cmp.saha.ernet.in}

\author{Srutarshi Pradhan}
 \email{spradhan@cmp.saha.ernet.in}

\author{Bikas K. Chakrabarti}
 \email{bikas@cmp.saha.ernet.in}

\affiliation{Theoretical Condensed Matter Physics Division,
 Saha Institute of Nuclear Physics,
 Sector - 1, Block - AF, Bidhannagar, Kolkata 700 064, India}

\date{February 19, 2003}

\begin{abstract}

We study the phase transition in a class of fiber bundle models
in which the fiber strengths are distributed randomly within a finite
interval and global load sharing is assumed. The dynamics is expressed
as recursion relations for the redistribution of the applied stress and
the evolution of the surviving fraction of fibers. We show that an
irreversible phase transition of second-order occurs, from a phase of
partial failure to a phase of total failure, when the initial applied
stress just exceeds a critical value.
The phase transition is characterised by static and
dynamic critical properties. We calculate exactly the critical value
of the initial stress for three models of this kind, each with a
different distribution of fiber strengths. We derive the exact
expressions for the order parameter, the susceptibility to changes in
the initial applied sress and the critical relaxation of the surviving
fraction of fibers for all the three models. The static and dynamic
critical exponents obtained from these expressions are found to be
universal.

\end{abstract}

\pacs{46.50.+a, 62.20.Mk, 64.60.Ht}

\maketitle

\section{Introduction}

\begin{figure}[htb]
\resizebox{16.5cm}{!}{\rotatebox{0}{\includegraphics{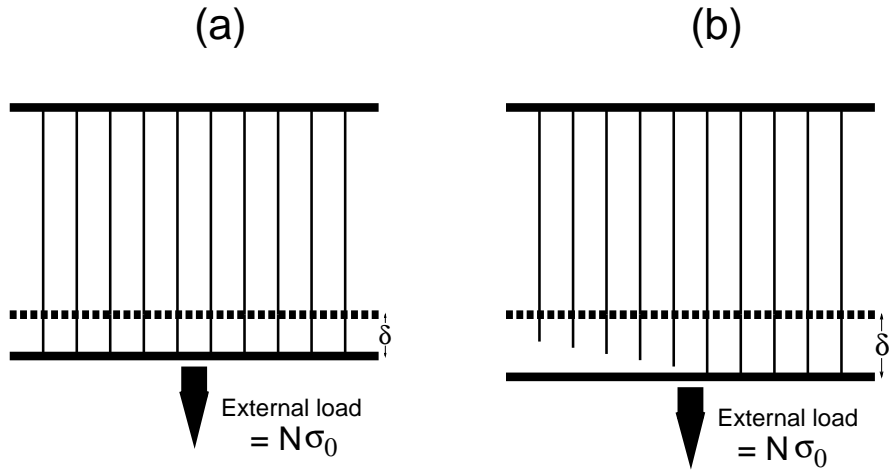}}}
\caption{Schematic diagram of a bundle of $N$ elastic
fibers with random strengths, attached in parallel to a fixed plate
at the top and a movable plate at the bottom. The fibers are shown as
vertical solid lines and the position of the plates by bold horizontal
lines. The figure shows the fibers arranged in increasing order of
strength from left to right; $\sigma_L$ and $\sigma_R$ are the strengths
of the weakest (extreme left) and the strongest (extreme right) fibers
respectively.
(a) For an initial applied stress $\sigma_0$,
$\sigma_0 \le \sigma_L$, the fibers only get stretched from their
relaxed position (shown by the bold broken line); the strain
$\delta$ measures the elastic deformation of the bundle.
(b) For $\sigma_L < \sigma_0 \le \sigma_R$, the fibers begin to
break causing a plastic deformation of the bundle (now given by the
strain $\delta '$); the figure schematically shows the strained
positions of the broken fibers at the time of their breaking in
order to give an impression of successive failures. When
$\sigma_0 > \sigma_R$, all fibers in the bundle break at once.}
\label{fig1}
\end{figure}

\indent Fiber bundle models describe the collective statics and
dynamics of failure in a set of fibers with random strengths under
the application of a stress (force per fiber)~\cite{Daniels1945,
Coleman1958}. A typical model of this kind is shown schematically
in Fig.~\ref{fig1}. These models are constructed for the purpose of
explaining the propagation of fractures in a loaded heterogeneous
material and to determine the conditions under which it breaks
completely~\cite{Herrmann1990, Chakrabarti1997}.
The latter requires the calculation of the strength of the
bundle from the strengths of its constituent fibers which,
by reasonable assumption, are drawn at random from a chosen
probability distribution~\cite{Daniels1945}. In some of the
models~\cite{Daniels1945, Sornette1989, Hemmer1992}
it is assumed that the load is always divided equally among all intact
fibers of the bundle (global load sharing) while in other models
~\cite{Harlow1981, Newman1995, Zhang1996, Zhang1999, Newman2001}
it is assumed that when a fiber breaks, the stress it was last
bearing gets distributed only among the fibers next to it
(local load sharing). The dynamics or propagation of fracture
in a fiber bundle has been characterised in two ways:
first, by the probability distribution of bursts of different sizes that
occur within the bundle as the stress is gradually increased till the
bundle breaks completely~\cite{Hemmer1992, Hansen1994, Kloster1997,
Hidalgo2001};
second, by the of lifetime of a fiber bundle with fatigue under an
applied stress~\cite{Coleman1958, Newman2001, Moral2001}.
It was suggested in~\cite{Zapperi1997} that the breakdown of a static
fiber bundle with global load sharing can be described as a
first-order phase transition, because the surviving fraction of fibers
has a discontinuity at the point of breakdown. However the susceptibility
to the applied stress was shown to diverge at the breakdown point
and for this reason it was later suggested that the transition is of
second-order~\cite{Andersen1997, daSilveira1998, daSilveira1999, Moreno2000}.

\indent In this paper we report on the universality of the phase
transition in a class of fiber bundle models. The main feature of these
models is a pair of dynamical recursion relations which, similar to the
formulation in~\cite{daSilveira1998, daSilveira1999}, expresses the
evolution of the fiber bundle under the application of a finite stress.
The strengths of the fibers are assigned randomly within a finite interval
of values which is true for real fiber bundles. We study three models of
this kind: in the first model the fiber strengths are distributed
with uniform density; in the second model the fiber strengths are
distributed with a linearly increasing density, which means that there
are more strong fibers than weak ones; in the third model the fiber
strengths are distributed with a linearly decreasing density so that
there are more weak fibers. From the expressions of the fixed points
of the dynamics we find that there is a critical initial value of
the applied stress in each of the three models: on exceeding this
critical value the fiber bundle undergoes an irreversible transition
from a phase of partial failure to a phase of total failure.
When the initial applied stress is less than or equal to
the critical value, only a finite fraction of the fibers breaks as the
bundle evolves to a state of mechanical equilibrium; this is the phase
of partial failure.
If, on the other hand, the initial applied stress is greater than the
critical value, mechanical equilibrium is never reached and the entire
fiber bundle eventually breaks down; this is the phase of total failure.
We define an order parameter which shows that the phase transition
is of second-order. As the initial applied stress approaches its critical
value from below, the order parameter is found to reduce to zero
continuously following a power-law while the susceptibility of the
surviving fraction of fibers to changes in the initial stress is found
to diverge, also by a power-law. We derive asymptotic solutions
of the dynamical recursion relations for the surviving fraction of
fibers at the critical values of the initial applied stress of each model.
These solutions show that the critical relaxation of the fiber bundle
toward the fixed point is a power-law decay. The critical exponents
in these power-laws are found to be universal,
i.e., independent of the distribution of fiber strengths in the bundle.

\section{Dynamics of the fiber bundle}

\indent We consider fiber bundle models with global load sharing
approximation, i.e., any force applied on a bundle is shared equally
by all the intact fibers in it. The strength of each fiber is determined
by a threshold value $\sigma_{\rm thresh}$ of the stress that it can
bear, beyond which the fiber breaks. The threshold stress of the fibers
in the bundle are distributed randomly with a normalised density
$\rho(\sigma_{\rm thresh})$ within a finite interval
$[\sigma_L, \sigma_R]$ where $\sigma_L$ and $\sigma_R$ are
respectively the strengths of the weakest and the strongest fiber in the
bundle:

\begin{equation}
\int_{\sigma_L}^{\sigma_R} \rho(\sigma_{\rm thresh}) \:
 {\rm d}\sigma_{\rm thresh} = 1.
\label{eq:normal-dens}
\end{equation}

\noindent The probability distribution of the threshold stress
is given by:

\begin{equation}
P(\sigma_{\rm thresh}) = \left \{ \begin{array}{l l}
 0, & ~0 \le \sigma_{\rm thresh} < \sigma_L \\
 \int_{\sigma_L}^{\sigma_{\rm thresh}} \rho(\sigma) {\rm d}\sigma,
 & \sigma_L \le \sigma_{\rm thresh} \le \sigma_R \\
 1, & \sigma_R < \sigma_{\rm thresh} .
                \end{array}
       \right .
\label{eq:prob-gen}
\end{equation}

\indent We study the breaking dynamics of the model under the application
of an initial stress $\sigma_0$ (for example, by attaching an external
load to the bottom plate in Fig.~\ref{fig1}) where the stress is defined as
the force exerted per fiber. The fibers whose strengths are less than
$\sigma_0$ break immediately. Following this initial rupture the applied
stress gets redistributed among the surviving fibers. Since the number of
fibers supporting the load has decreased the redistributed stress is greater
than the initial stress and this causes further breaking of fibers.
The process continues till a state of mechanical equilibrium is reached,
where the surviving fibers have strengths greater than the redistributed
stress, or till all fibers in the bundle are broken. The state of
mechanical equilibrium, if it exists, appears as a fixed point of the
model under the assigned dynamics.

\indent The nature of the breaking dynamics allows it to be represented
as a recursion relation operating in discrete time-steps
~\cite{Phoenix1979, daSilveira1998, daSilveira1999}.
If $U_t$ is the fraction of fibers in the initial bundle that survive
after time-step $t$, the redistributed stress due to global load
sharing after $t$ time-steps is

\begin{equation}
\sigma_t = {\sigma_0 \over U_t} .
\label{eq:stress-defn}
\end{equation}

\noindent After $t+1$ time-steps the surviving fraction of fibers becomes

\begin{equation}
U_{t+1} = 1 - P(\sigma_t).
\label{eq:frac-defn}
\end{equation}

\noindent In a real bundle comprising of a finite number of fibers
there will be fluctuations in the local density of fiber strengths
as well as the load sharing. Such fluctuatins are ignored in the
contruction of Eq.~(\ref{eq:stress-defn}) and Eq.~(\ref{eq:frac-defn}).
It follows from Eq.~(\ref{eq:stress-defn}) and Eq.~(\ref{eq:frac-defn})
that the quantities $\sigma_t$ and $U_t$ evolve by the recurrences

\begin{equation}
\sigma_{t+1} = {\sigma_0 \over {1 - P(\sigma_t)}}
\label{eq:stressrecur-gen}
\end{equation}

\noindent and
\begin{equation}
U_{t+1} = 1 - P(\sigma_0 / U_t), \hspace{1.0cm} U_0 = 1,
\label{eq:fracrecur-gen}
\end{equation}

\noindent which formally define the dynamics of this class of models.
With the probability distribution Eq.~(\ref{eq:prob-gen}), the recursion
relation (\ref{eq:fracrecur-gen}) clearly shows that none of the
fibers break for an initial stress $\sigma_0 \le \sigma_L$, while on
application of $\sigma_0 > \sigma_R$, all fibers in the bundle break
simultaneously.

\indent The fixed points~\cite{footnote1}
of the model, $\sigma^*$ and $U^*$, are determined by the relations:

\begin{equation}
\sigma^*[1 - P(\sigma^*)] = \sigma_0
\label{eq:stressfix-gen}
\end{equation}

\noindent and
\begin{equation}
U^* + P(\sigma_0/U^*) = 1.
\label{eq:fracfix-gen}
\end{equation}

\indent Though the quantities $\sigma_t$ and $U_t$ evolve in time
till they reach their fixed point values, Eq.~(\ref{eq:stress-defn})
shows that their product $\sigma_t U_t$ is a constant of motion,
always equal to the initial value $\sigma_0$. All static and dynamic
properties of the models are consequences of this invariance.

\section{Critical properties for uniform density of fiber strengths}

\begin{figure}[htb]
\resizebox{13.0cm}{!}{\rotatebox{0}{\includegraphics{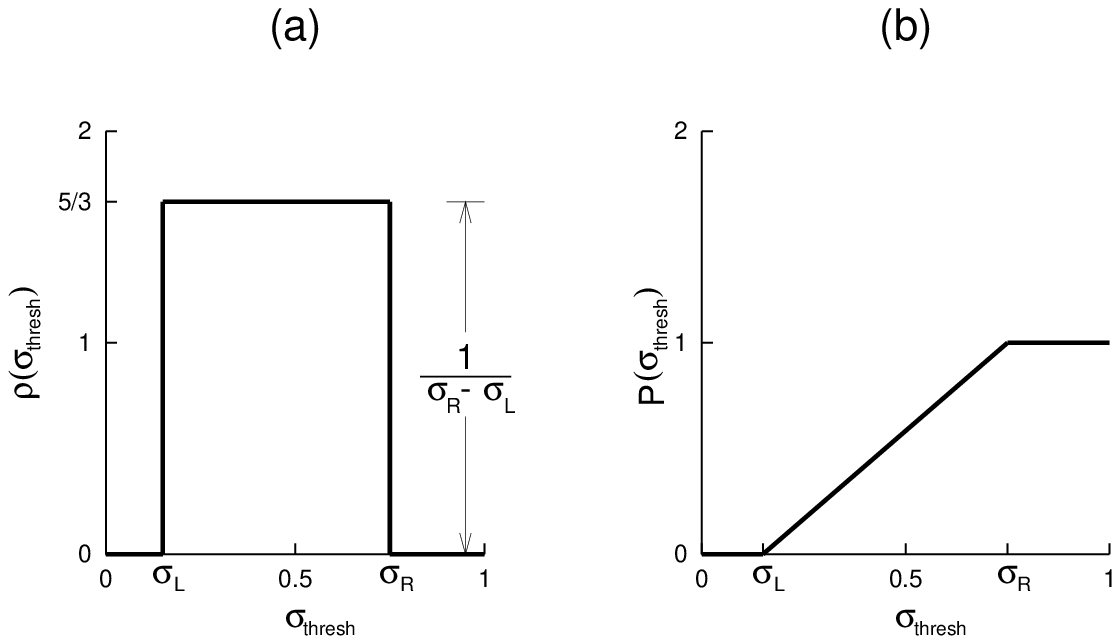}}}
\caption{(a) The density function $\rho$ and (b) the
probability distribution $P$ of random fiber strengths
$\sigma_{\rm thresh}$ distributed with uniform density in the
interval $[\sigma_L, \sigma_R]$. In the particular instance shown
in the figure $\sigma_L = 0.15$ and $\sigma_R = 0.75$.}
\label{fig2}
\end{figure}

\indent We consider first the case where the random strengths
$\sigma_{\rm thresh}$ of the fibers are distributed with uniform
density in the interval $[\sigma_L, \sigma_R]$. The normalised density
function is:

\begin{equation}
\rho(\sigma_{\rm thresh}) = \left \{ \begin{array}{l l}
 0, & ~0 \le \sigma_{\rm thresh} < \sigma_L \\
 {1 \over {\sigma_R - \sigma_L}}, & \sigma_L \le \sigma_{\rm thresh}
    \le \sigma_R \\
 0, & \sigma_R < \sigma_{\rm thresh}
                \end{array}
       \right .
\label{eq:dens-uni}
\end{equation}

\noindent and the probability distribution, by the definition
in Eq.~(\ref{eq:prob-gen}), is given by (Fig.~\ref{fig2}):

\begin{equation}
P(\sigma_{\rm thresh}) = \left \{ \begin{array}{l l}
 0, & ~0 \le \sigma_{\rm thresh} < \sigma_L \\
 {{\sigma_{\rm thresh} - \sigma_L} \over {\sigma_R - \sigma_L}},
    & \sigma_L \le \sigma_{\rm thresh} \le \sigma_R \\
 1, & \sigma_R < \sigma_{\rm thresh} .
                \end{array}
       \right .
\label{eq:prob-uni}
\end{equation}

\noindent We consider only $\sigma_R > \sigma_L$ as the case of
$\sigma_R = \sigma_L$ is trivial.

\indent With this particular choice of the distribution of fiber
strengths the recursion relations (\ref{eq:stressrecur-gen})
and (\ref{eq:fracrecur-gen}), for an initial stress
$\sigma_L \le \sigma_0 \le \sigma_R$, appear as:

\begin{equation}
\sigma_{t+1} =  \sigma_0 \left ( {{\sigma_R -  \sigma_L}
                                   \over {\sigma_R - \sigma_t}} \right )
\label{eq:stressrecur-uni}
\end{equation}

\noindent and
\begin{equation}
U_{t+1} =  {1 \over {\sigma_R - \sigma_L}}
           \left ( \sigma_R -  {\sigma_0 \over U_t} \right ),
 \hspace{1.0cm} U_0 = 1.
\label{eq:fracrecur-uni}
\end{equation}

\noindent The fixed point equations, obtained from
Eq.~(\ref{eq:stressfix-gen}) and Eq.~(\ref{eq:fracfix-gen}),
appear in quadratic form:

\begin {equation}
\left ( \sigma^* \right )^2 - \sigma_R \sigma^*
 + \sigma_0 \left (\sigma_R - \sigma_L \right ) = 0
\label{eq:stressfix-uni}
\end {equation}

\noindent and
\begin {equation}
\left ( U^* \right )^2
 - \left ( {\sigma_R \over {\sigma_R - \sigma_L}} \right ) U^*
 + {\sigma_0 \over {\sigma_R - \sigma_L}} = 0.
\label{eq:fracfix-uni}
\end {equation}

\noindent Consequently each of the recurrences, (\ref{eq:stressrecur-uni})
and (\ref{eq:fracrecur-uni}), have two fixed points:

\begin{equation}
\sigma^*_{1,2} =  {\sigma_R \over 2} \pm (\sigma_R - \sigma_L)^{1/2}
            \left [ {\sigma_R^2 \over {4(\sigma_R - \sigma_L)}}
             - \sigma_0 \right ]^{1/2},
\label{eq:stressfix-uni12}
\end{equation}

\begin{equation}
U^*_{1,2} =  {\sigma_R \over 2(\sigma_R - \sigma_L)} \pm
       {1 \over (\sigma_R - \sigma_L)^{1/2}}
            \left [ {\sigma_R^2 \over {4(\sigma_R - \sigma_L)}}
             - \sigma_0 \right ]^{1/2}.
\label{eq:fracfix-uni12}
\end{equation}

\begin{figure}[htb]
\resizebox{13.0cm}{!}{\rotatebox{0}{\includegraphics{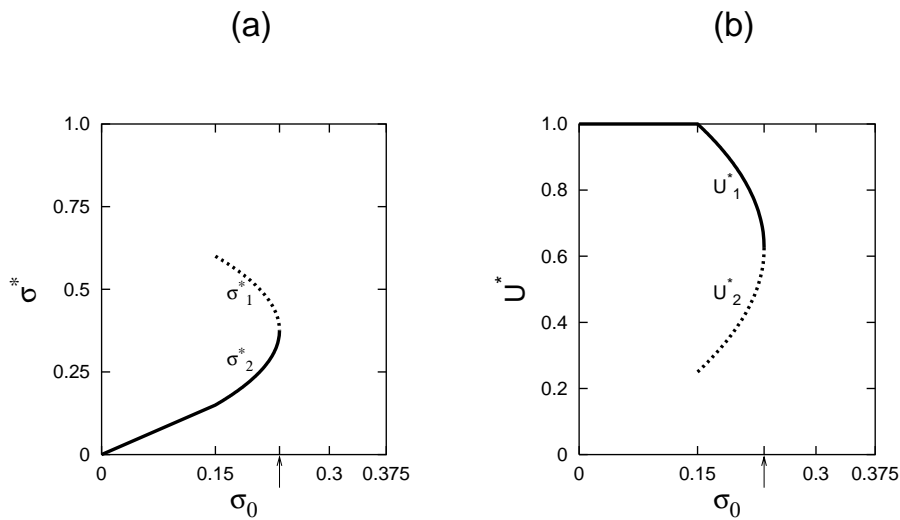}}}
\caption{\footnotesize The fixed points of (a) the redistributed stress,
and (b) the surviving fraction of fibers for the particular
probability distribution of fiber strengths shown in Fig.~\ref{fig2}.
In each part of the figure the curve for the stable fixed points is shown
by a bold solid line and that for the unstable fixed points are shown
by a bold broken line. As in Fig.~\ref{fig2} we have $\sigma_L = 0.15$
and $\sigma_R = 0.75$, so that $\sigma_0^{\rm crit} = 0.234375$;
the position of the critical point is marked by an arrowhead.
For $\sigma_0 \le \sigma_L$ the fixed points are trivial: since there
are no broken fibers $\sigma^* = \sigma_0$ and $U^* = U_0 = 1$.}
\label{fig3}
\end{figure}

\noindent The subscripts $1$ and $2$ stand for the expressions containing
the plus and the minus sign respectively. While $\sigma_2^*$ and $U_1^*$
are stable fixed points, $\sigma_1^*$ and $U_2^*$ are unstable
(Fig.~\ref{fig3}). It is clear that the fixed points for the redistributed
stress and the surviving fraction of fibers are related by:

\begin{equation}
U^*_{1,2} = {{\sigma_R - \sigma^*_{2,1}} \over {\sigma_R - \sigma_L}}.
\label{eq:stress-frac-fix}
\end{equation}

\indent The quadratic equations (\ref{eq:stressfix-uni}) and
(\ref{eq:fracfix-uni}) show that the initial applied stress
$\sigma_0$ has a critical value:

\begin{equation}
\sigma_0^{\rm crit} =  {\sigma_R^2 \over {4(\sigma_R - \sigma_L})},
\label{eq:stresscrit-uni}
\end{equation}

\noindent at which their discriminants become zero and only a stable
fixed point exists. Since both $\sigma$ and $U$ have physical meanings, a
state of mechanical equilibrium exists if the quantities $\sigma_2^*$
and $U_1^*$ are positive real-valued; this happens only when
$\sigma_0 \le \sigma_0^{\rm crit}$.
For $\sigma_0 > \sigma_0^{\rm crit}$, there are no real-valued fixed
points -- the dynamics will continue till all fibers in the bundle
are broken. Therefore a transition ocurs just above
$\sigma_0^{\rm crit}$ from a phase of partial failure of the bundle
in equilibrium to a phase of total failure.
The order parameter ${\cal O}$ for this phase transition
is defined in terms of the stable fixed point for the surviving
fraction of fibers:

\begin{equation}
{\cal O} \equiv U^*_1 - U^*_{\rm 1-crit}, \hspace{1.0cm}
 \sigma_L \le \sigma_0 \le \sigma_0^{\rm crit}
\label{eq:order-uni-defn}
\end{equation}

\noindent where
\begin{equation}
U^*_{\rm 1-crit} =  {\sigma_R \over 2(\sigma_R - \sigma_L)}
\label{eq:fracfix-uni-crit}
\end{equation}

\noindent is the value of $U^*_1$ under the critical initial stress.
As Eq.~(\ref{eq:fracfix-uni12}) shows, the order parameter
goes to zero continuously by a power law as $\sigma_0$ approaches
its critical value from below:

\begin{equation}
{\cal O} = \left ( {\sigma_0^{\rm crit} - \sigma_0 \over
\sigma_R - \sigma_L} \right )^{1/2},
 \hspace{1.0cm} \sigma_L \leq \sigma_0 \leq \sigma_0^{\rm crit}.
\label{eq:order-uni-crit}
\end{equation}

\indent It is obvious that the critical value must have the
lower bound:

\begin{equation}
\sigma_0^{\rm crit} \ge \sigma_L,
\label{eq:stresscrit-uni-low}
\end{equation}

\noindent which, for the expression given in
Eq.~(\ref{eq:stresscrit-uni}), requires that

\begin{equation}
\sigma_R \ge 2 \sigma_L.
\label{eq:weak-strong-relate-uni}
\end{equation}

\noindent The above condition (\ref{eq:weak-strong-relate-uni}) in
turn imposes an upper bound on the critical value of the initial stress:

\begin{equation}
\sigma_0^{\rm crit} \le {\sigma_R \over 2}.
\label{eq:stresscrit-uni-up}
\end{equation}

\indent Static critical behaviour of the fiber bundle is observed in
the susceptibility of the fixed point of the surviving fraction of
fibers to changes in the initial stress $\sigma_0$. From
Eq.~(\ref{eq:fracfix-uni12}) we see that the susceptibility diverges
in the form of a power law as the initial applied stress approaches
its critical value from below:

\begin{equation}
\chi = \left | {{\rm d} U^*_1 \over {\rm d}\sigma_0} \right |
     \propto \left (\sigma_0^{\rm crit} - \sigma_0 \right ) ^{-1/2},
 \hspace{1.0cm} \sigma_L \leq \sigma_0 \leq \sigma_0^{\rm crit}.
\label{eq:suscept-uni-crit}
\end{equation}

\indent Dynamical critical behaviour is observed in the process of
relaxation of the fiber bundle to a fixed point.
At the critical point $\sigma_0 = \sigma_0^{\rm crit}$
the evolution of the surviving fraction of fibers by
(\ref{eq:fracrecur-uni}) is reduced to the following recurrence:

\begin{equation}
U_{t+1} =  {\sigma_R \over {\sigma_R - \sigma_L}}
            \left [ 1 - {\sigma_R \over {4(\sigma_R - \sigma_L)}}
            {1 \over U_t} \right ], \hspace{1.0cm} U_0 = 1,
\label{eq:fracrecur-uni-crit}
\end{equation}

\noindent and its only fixed point, a stable one, is given by
Eq.~(\ref{eq:fracfix-uni-crit}).

\noindent The recurrence (\ref{eq:fracrecur-uni-crit}) has
a closed-form solution:

\begin{equation}
U_t - U^*_{\rm 1-crit} = {U^*_{\rm 1-crit} (1 - U^*_{\rm 1-crit})
                           \over {1 + (1 - U^*_{\rm 1-crit}) (t - 1)}}.
\label{eq:frac-uni-crit}
\end{equation}

\noindent Thus the asymptotic behaviour of the surviving fraction of fibers
is a power law decay -- a critically slow relaxation to the fixed point:

\begin{eqnarray}
U_t - U^*_{\rm 1-crit} & \sim & {U^*_{\rm 1-crit} \over t},
 \hspace{1.0cm} t \to \infty \nonumber \\
~ & \sim & {1 \over 2} \left ( {\sigma_R \over \sigma_R - \sigma_L} \right )
           {1 \over t}.
\label{eq:frac-uni-crit-dyn}
\end{eqnarray}

\indent A special case of this model, with $\sigma_L = 0$ and
$\sigma_R = 1$, was studied in~\cite{Pradhan2002}. The critical
properties obtained in~\cite{Pradhan2002} can now be derived
easily from the general results of this section.

\section{Critical properties for linearly increasing density
of fiber strengths}

\begin{figure}[htb]
\resizebox{13.0cm}{!}{\rotatebox{0}{\includegraphics{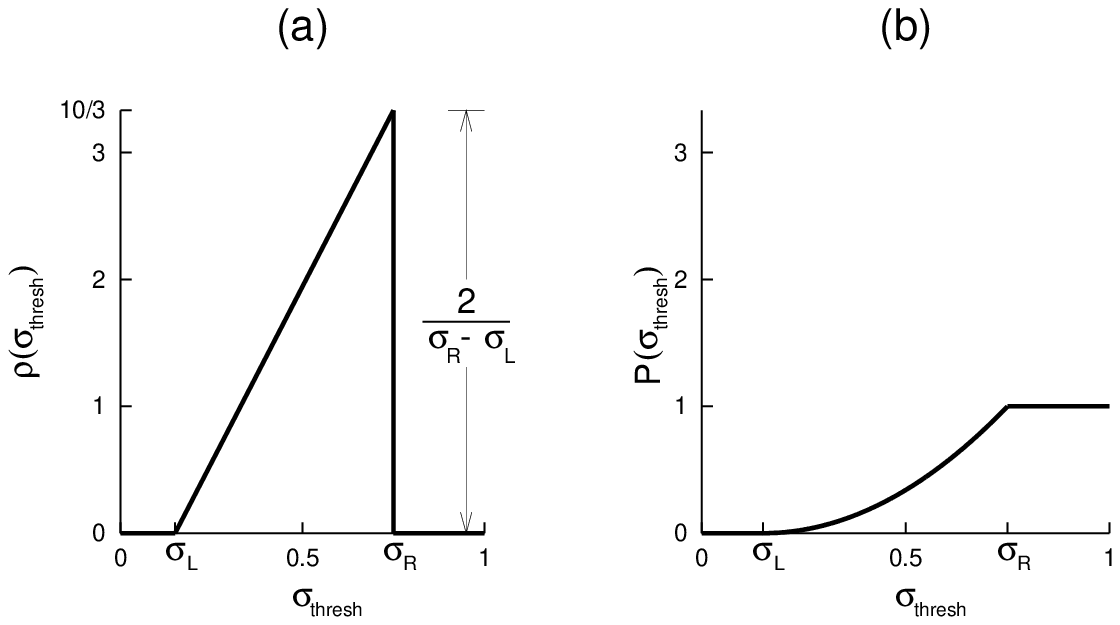}}}
\caption{(a) The density function $\rho$ and (b) the
probability distribution $P$ of random fiber strengths
$\sigma_{\rm thresh}$ distributed with linearly increasing density
in the interval $[\sigma_L, \sigma_R]$. In the particular instance shown
in the figure $\sigma_L = 0.15$ and $\sigma_R = 0.75$.}
\label{fig4}
\end{figure}

\indent We consider next the case where the random strengths
$\sigma_{\rm thresh}$ of the fibers are distributed with linearly
increasing density in the interval $[\sigma_L, \sigma_R]$,
$\sigma_R > \sigma_L$. The normalised density function and the
probability distribution of the fiber strengths are given by
(illustrated in Fig.~\ref{fig4}):

\begin{equation}
\rho(\sigma_{\rm thresh}) = \left \{ \begin{array}{l l}
 0, & ~0 \le \sigma_{\rm thresh} < \sigma_L \\
 {{2(\sigma_{\rm thresh} - \sigma_L)} \over {(\sigma_R - \sigma_L)^2}},
    & \sigma_L \le \sigma_{\rm thresh} \le \sigma_R \\
 0, & \sigma_R < \sigma_{\rm thresh}
                \end{array}
       \right .
\label{eq:dens-lin}
\end{equation}

\noindent and
\begin{equation}
P(\sigma_{\rm thresh}) = \left \{ \begin{array}{l l}
 0, & ~0 \le \sigma_{\rm thresh} < \sigma_L \\
 \left ({{\sigma_{\rm thresh} - \sigma_L} \over {\sigma_R - \sigma_L}}
 \right )^2,
    & \sigma_L \le \sigma_{\rm thresh} \le \sigma_R \\
 1, & \sigma_R < \sigma_{\rm thresh} .
                \end{array}
       \right .
\label{eq:prob-lin}
\end{equation}

\indent Here we introduce the transformed quantities:

\begin{equation}
\Gamma_0 = {\sigma_0 \over {\sigma_R - \sigma_L}}, \hspace{1.0cm}
\Gamma_L = {\sigma_L \over {\sigma_R - \sigma_L}}, \hspace{1.0cm}
\Gamma_t = {\sigma_t \over {\sigma_R - \sigma_L}}.
\label{eq:trans-stress}
\end{equation}

\indent For an initial stress $\sigma_L \leq \sigma_0 \leq \sigma_R$
(or, $\Gamma_L \leq \Gamma_0 \leq \Gamma_L+1$) along with the distribution
of fiber strengths given by Eq.~(\ref{eq:prob-lin}), the recursion
relations (\ref{eq:stressrecur-gen}) and (\ref{eq:fracrecur-gen})
appear as:

\begin{equation}
\Gamma_{t+1} = {\Gamma_0 \over {1 - (\Gamma_t - \Gamma_L)^2}}
\label{eq:stressrecur-lin}
\end{equation}

\noindent and
\begin{equation}
U_{t+1} = 1 - \left ({\Gamma_0 \over U_t} - \Gamma_L \right )^2 ,
 \hspace{1.0cm} U_0 = 1.
\label{eq:fracrecur-lin}
\end{equation}

\noindent The fixed point equations, (\ref{eq:stressfix-gen}) and
(\ref{eq:fracfix-gen}), now assume cubic form:

\begin{equation}
\left (\Gamma^* \right )^3 - 2 \Gamma_L \left (\Gamma^* \right )^2
 + \left (\Gamma_L^2 - 1 \right ) \Gamma^* + \Gamma_0 = 0
\label{eq:stressfix-lin}
\end{equation}

\noindent where $\Gamma^* = \sigma^* / (\sigma_R - \sigma_L)$, and

\begin{equation}
\left (U^* \right )^3 + \left ( \Gamma_L^2 - 1 \right ) \left (U^* \right )^2
 - \left (2 \Gamma_L \Gamma_0 \right ) U^* + \Gamma_0^2 = 0.
\label{eq:fracfix-lin}
\end{equation}

\noindent Consequently each of the recurrences (\ref{eq:stressrecur-lin})
and (\ref{eq:fracrecur-lin}) have three fixed points -- only one in
each case is found to be stable. For the redistributed stress the fixed
points are:

\begin{eqnarray}
\Gamma^*_1 & = & \frac{2}{3} \Gamma_L + 2 K \cos\frac{\Phi}{3},
\label{eq:stressfix-lin1} \\
\Gamma^*_2 & = & \frac{2}{3} \Gamma_L - K \cos\frac{\Phi}{3}
 + \sqrt{3} K \sin\frac{\Phi}{3},
\label{eq:stressfix-lin2} \\
\Gamma^*_3 & = & \frac{2}{3} \Gamma_L - K \cos\frac{\Phi}{3}
 - \sqrt{3} K \sin\frac{\Phi}{3},
\label{eq:stressfix-lin3}
\end{eqnarray}

\noindent where
\begin{equation}
K = \frac{1}{3} \sqrt{3 + \Gamma_L^2}
\label{eq:K-defn}
\end{equation}

\noindent and
\begin{equation}
\cos\Phi = {\Gamma_L \left ( 9 - \Gamma_L^2 \right ) - 27 \Gamma_0 / 2
            \over \left ( 3 + \Gamma_L^2 \right )^{3/2}}.
\label{eq:Phi-defn}
\end{equation}

\noindent Similarly, for the surviving fraction of fibers the fixed
points are:

\begin{eqnarray}
U^*_1 & = & {1 - \Gamma_L^2 \over 3} + 2 J \cos\frac{\Theta}{3},
\label{eq:fracfix-lin1} \\
U^*_2 & = & {1 - \Gamma_L^2 \over 3} - J \cos\frac{\Theta}{3}
 + \sqrt{3} J \sin\frac{\Theta}{3},
\label{eq:fracfix-lin2} \\
U^*_3 & = & {1 - \Gamma_L^2 \over 3} - J \cos\frac{\Theta}{3}
 - \sqrt{3} J \sin\frac{\Theta}{3},
\label{eq:fracfix-lin3}
\end{eqnarray}

\noindent where
\begin{equation}
J = \frac{1}{3} \sqrt{(\Gamma_L^2 - 1)^2 + 6 \Gamma_L \Gamma_0}
\label{eq:J-defn}
\end{equation}

\noindent and
\begin{equation}
\cos\Theta = {\left ( 1 - \Gamma_L^2 \right )
 \left [ \left ( \Gamma_L^2 - 1 \right )^2 + 9 \Gamma_L \Gamma_0 \right ]
 - 27 \Gamma_0^2 / 2
 \over \left [ \left ( \Gamma_L^2 - 1 \right )^2
               + 6 \Gamma_L \Gamma_0 \right ]^{3/2}}.
\label{eq:Theta-defn}
\end{equation}

\begin{figure}[htb]
\resizebox{13.0cm}{!}{\rotatebox{0}{\includegraphics{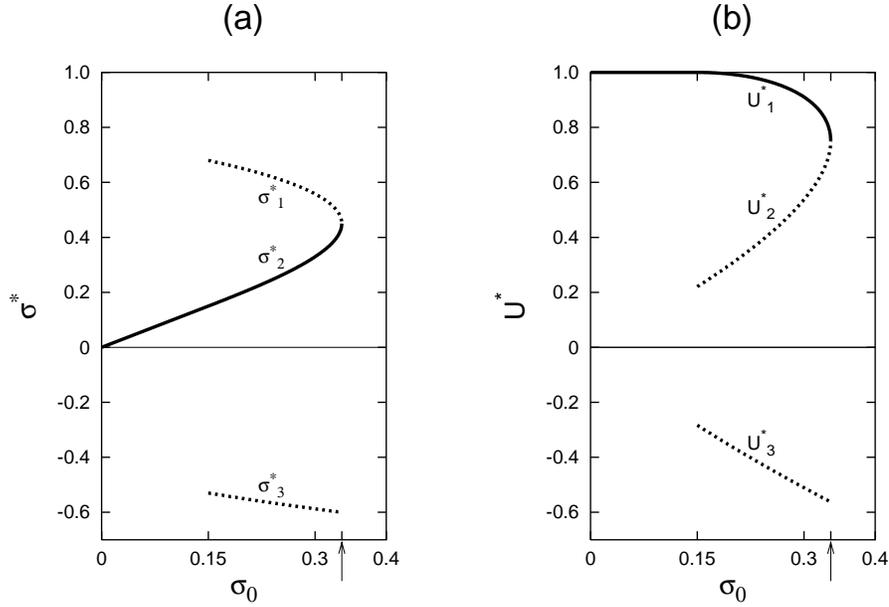}}}
\caption{The fixed points of (a) the redistributed stress,
and (b) the surviving fraction of fibers for the probability distribution
of fiber strengths shown in Fig.~\ref{fig4}.
In each part of the figure the curve for the stable fixed points is shown
by a bold solid line and those for the unstable fixed points are shown
by bold broken lines. We have $\sigma_L = 0.15$ and $\sigma_R = 0.75$,
so that $\sigma_0^{\rm crit} = 0.3375$; the position of
the critical point is marked by an arrowhead. As in the previous case,
for $\sigma_0 \le \sigma_L$ the fixed points are trivial: since there
are no broken fibers $\sigma^* = \sigma_0$ and $U^* = U_0 = 1$.}
\label{fig5}
\end{figure}

\noindent Of these fixed points $\Gamma^*_2$ and $U^*_1$ are stable
whereas $\Gamma^*_1$, $\Gamma^*_3$ and $U^*_2$, $U^*_3$ are unstable
(Fig.~\ref{fig5}).

\indent Similar to the case in the previous section the
discriminants of the cubic equations (\ref{eq:stressfix-lin}) and
(\ref{eq:fracfix-lin}) become zero at a critical value
$\sigma_0^{\rm crit}$ (or, $\Gamma_0^{\rm crit}$) of the initial
applied stress:

\begin{eqnarray}
\Gamma_0^{\rm crit} & = & {\sigma_0^{\rm crit}
                               \over {\sigma_R - \sigma_L}} \nonumber \\
~ & = & {2 \over 27} \left [ \Gamma_L \left ( 9 - \Gamma_L^2 \right )
                             + \left ( 3 + \Gamma_L^2 \right )^{3/2} \right ]
\label{eq:stresscrit-lin}
\end{eqnarray}

\noindent and then each of the quantities $\Gamma$ and $U$ have one
stable and one unstable fixed point. As before the critical point has
the trivial lower bound:

\begin{equation}
\sigma_0^{\rm crit} \geq \sigma_L.
\label{eq:stresscrit-lin-low}
\end{equation}

\noindent The expression of $\Gamma_0^{\rm crit}$ in
Eq.~(\ref{eq:stresscrit-lin}) shows that it approaches the lower bound
as $\Gamma_L \to \infty$ which happens for finite values of $\sigma_L$
and $\sigma_R$ when $(\sigma_R - \sigma_L) \to 0$. It follows that
the upper bound for the critical point is also trivial:

\begin{equation}
\sigma_0^{\rm crit} \leq \sigma_R,
\label{eq:stresscrit-lin-up}
\end{equation}

\noindent which is different from the condition
(\ref{eq:stresscrit-uni-up}) for the case of uniform distribution.

\noindent At the critical point we get from Eq.~(\ref{eq:Phi-defn})
and Eq.~(\ref{eq:Theta-defn}):

\begin{equation}
\cos \Phi_{\rm crit} = \cos \Theta_{\rm crit} = -1
\label{eq:Phi-Theta-crit1}
\end{equation}

\noindent or,
\begin{equation}
\Phi_{\rm crit} = \Theta_{\rm crit} = \pi.
\label{eq:Phi-Theta-crit2}
\end{equation}

\indent The stable fixed points $\Gamma^*_2$ and $U_1^*$ are
positive real-valued
when $\Gamma_0 \leq \Gamma_0^{\rm crit}$; thus the fiber bundle
always reaches a state of mechanical equilibrium after partial failure
under an initial applied stress $\sigma_0 \leq \sigma_0^{\rm crit}$.
For $\sigma_0 > \sigma_0^{\rm crit}$ (or, $\Gamma_0 >
\Gamma_0^{\rm crit}$), $\Gamma^*_2$ and $U_1^*$ are no longer
real-valued and the entire fiber bundle eventually breaks down.
The transition from the phase of partial failure to the phase of
total failure takes place when $\sigma_0$ just exceeds
$\sigma_0^{\rm crit}$ and the order parameter for this phase
transition is defined as in Eq.~(\ref{eq:order-uni-defn}):

\begin{equation}
{\cal O} \equiv U^*_1 - U^*_{\rm 1-crit}.
\label{eq:order-lin-defn}
\end{equation}

\noindent Close to the critical point but below it, we can write,
from Eq.~(\ref{eq:Theta-defn}) and Eq.~(\ref{eq:Phi-Theta-crit2}),
that:

\begin{eqnarray}
\pi - \Theta & \simeq & \sin \Theta \nonumber \\
~ & \simeq  & {3 \sqrt{3} \: \Gamma_0^{\rm crit} (3 + \Gamma_L^2)^{3/4}
 (\Gamma_0^{\rm crit} -\Gamma_0)^{1/2} \over
 [(\Gamma_L^2 - 1)^2 + 6 \Gamma_L \Gamma_0^{\rm crit}]^{3/2}}
\label{eq:Theta-nearcrit}
\end{eqnarray}

\noindent and the expressions for the fixed points in
Eq.~(\ref{eq:fracfix-lin1}) and Eq.~(\ref{eq:fracfix-lin2}) reduce
to the forms:

\begin{equation}
U^*_1 \simeq U^*_{\rm 1-crit} +
 {\Gamma_0^{\rm crit} (3 + \Gamma_L^2)^{3/4}
 \over (\Gamma_L^2 - 1)^2 + 6 \Gamma_L \Gamma_0^{\rm crit}}
 \left ( \Gamma_0^{\rm crit} - \Gamma_0 \right )^{1/2}
\label{eq:fracfix-lin1-nearcrit}
\end{equation}

\noindent and
\begin{equation}
U^*_2 \simeq U^*_{\rm 2-crit} -
 {\Gamma_0^{\rm crit} (3 + \Gamma_L^2)^{3/4}
 \over (\Gamma_L^2 - 1)^2 + 6 \Gamma_L \Gamma_0^{\rm crit}}
 \left ( \Gamma_0^{\rm crit} - \Gamma_0 \right )^{1/2},
\label{eq:fracfix-lin2-nearcrit}
\end{equation}

\noindent where
\begin{equation}
U^*_{1-{\rm crit}} = U^*_{2-{\rm crit}} = {1 - \Gamma_L^2 \over 3} +
 {1 \over 3} \sqrt{(\Gamma_L^2 - 1)^2 + 6 \Gamma_L \Gamma_0^{\rm crit}}
\label{eq:fracfix-lin12-crit}
\end{equation}

\noindent is the stable fixed point value of the surviving fraction
of fibers under the critical initial stress $\sigma_0^{\rm crit}$.
Therefore, following the definition of the order parameter in
Eq.~(\ref{eq:order-lin-defn}) we get from the above equation:

\begin{equation}
{\cal O} = {\Gamma_0^{\rm crit} (3 + \Gamma_L^2)^{3/4}
 \over (\Gamma_L^2 - 1)^2 + 6 \Gamma_L \Gamma_0^{\rm crit}}
 \left ( \Gamma_0^{\rm crit} - \Gamma_0 \right )^{1/2},
 \hspace{1.0cm} \Gamma_0 \to \Gamma_0^{\rm crit}-.
\label{eq:order-lin-crit}
\end{equation}

\noindent On replacing the transformed variable $\Gamma_0$ by the original
$\sigma_0$, Eq.~(\ref{eq:order-lin-crit}) shows that the order parameter
goes to zero continuously following the same power-law as in
Eq.~(\ref{eq:order-uni-crit})
for the previous case when $\sigma_0$ approaches its critical
value from below.

\indent Similarly the susceptibility diverges by the same power-law
as in Eq.~(\ref{eq:suscept-uni-crit}) on approaching the critical point
from below:

\begin{equation}
\chi = \left | {{\rm d} U^*_1 \over {\rm d}\sigma_0} \right |
     \propto \left (\Gamma_0^{\rm crit} - \Gamma_0 \right ) ^{-1/2},
 \hspace{1.0cm} \Gamma_0 \to \Gamma_0^{\rm crit}-.
\label{eq:suscept-lin-crit}
\end{equation}

\indent The critical dynamics of the fiber bundle is given by the
asymptotic closed form solution of the recurrence
(\ref{eq:fracrecur-lin}) for $\Gamma_0 = \Gamma^{\rm crit}_0$:

\begin{equation}
U_t - U^*_{\rm 1-crit} \sim \left [
 {\left ( U^*_{\rm 1-crit} \right ) ^4
 \over 3 \left ( \Gamma^{\rm crit}_0 \right ) ^2 - 2 \Gamma_L
 \Gamma_0^{\rm crit} U^*_{\rm 1-crit}} \right ] {1 \over t},
 \hspace{1.0cm} t \to \infty,
\label{eq:frac-lin-crit-dyn}
\end{equation}

\noindent where $\Gamma_0^{\rm crit}$ and $U^*_{\rm 1-crit}$
are given in Eq.~(\ref{eq:stresscrit-lin}) and
Eq.~(\ref{eq:fracfix-lin12-crit}) respectively.
This shows that the asymptotic relaxation of the surviving
fraction of fibers to its stable fixed point under the critical initial
stress has the same inverse of time form as found in the case of
uniform density of fiber strengths [Eq.~(\ref{eq:frac-uni-crit-dyn})].

\section{Critical properties for linearly decreasing density
of fiber strengths}

\begin{figure}[htb]
\resizebox{13.0cm}{!}{\rotatebox{0}{\includegraphics{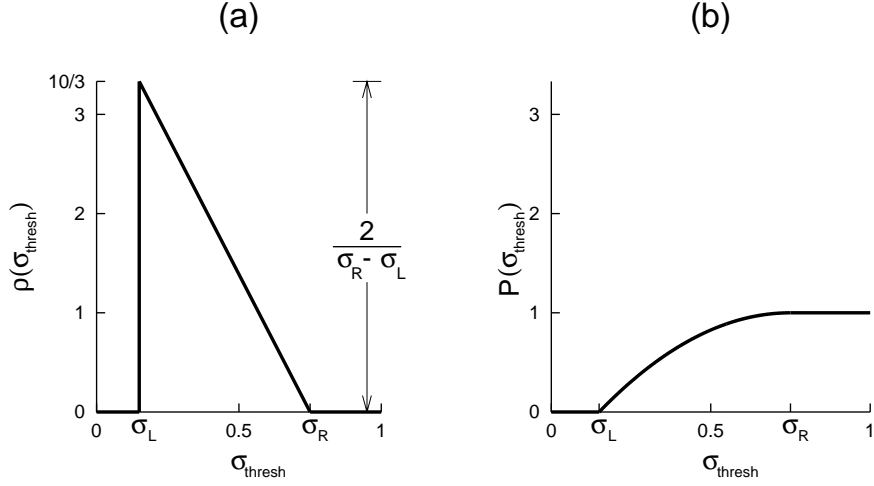}}}
\caption{(a) The density function $\rho$ and (b) the
probability distribution $P$ of random fiber strengths
$\sigma_{\rm thresh}$ distributed with linearly decreasing density
in the interval $[\sigma_L, \sigma_R]$. Similar to the cases shown in
Fig.~\ref{fig2} and Fig.~\ref{fig4} we have $\sigma_L = 0.15$ and
$\sigma_R = 0.75$ in this example also.}
\label{fig6}
\end{figure}

\indent Contrary to the case of the previoius section we now consider
a fiber bundle where the random threshold values are distributed
with a linearly decreasing density in the interval
$[\sigma_L, \sigma_R]$, $\sigma_R > \sigma_L$.
Instead of Eq.~(\ref{eq:dens-lin}) and Eq.~(\ref{eq:prob-lin})
we now have the following normalised density function and
probability distribution (illustrated in Fig.~\ref{fig6}):

\begin{equation}
\rho(\sigma_{\rm thresh}) = \left \{ \begin{array}{l c}
 0, & 0 \le \sigma_{\rm thresh} < \sigma_L \\
 {{2(\sigma_R - \sigma_{\rm thresh})} \over {(\sigma_R - \sigma_L)^2}},
    & \sigma_L \le \sigma_{\rm thresh} \le \sigma_R \\
 0, & \sigma_R < \sigma_{\rm thresh}
                \end{array}
       \right .
\label{eq:dens-lin'}
\end{equation}

\noindent and
\begin{equation}
P(\sigma_{\rm thresh}) = \left \{ \begin{array}{l c}
 0, & 0 \le \sigma_{\rm thresh} < \sigma_L \\
 1 - \left ({{\sigma_R - \sigma_{\rm thresh}} \over {\sigma_R - \sigma_L}}
 \right )^2,
    & \sigma_L \le \sigma_{\rm thresh} \le \sigma_R \\
 1, & \sigma_R < \sigma_{\rm thresh}
                \end{array}
       \right .
\label{eq:prob-lin'}
\end{equation}

\indent With the transformed quantities defined in Eq.~(\ref{eq:trans-stress})
the recurrences (\ref{eq:stressrecur-gen}) and (\ref{eq:fracrecur-gen})
for $\sigma_L \leq \sigma_0 \leq \sigma_R$ appear as:

\begin{equation}
\Gamma_{t+1} = {\Gamma_0 \over \left ( 1 + \Gamma_L - \Gamma_t \right )^2}
\label{eq:stressrecur-lin'}
\end{equation}

\noindent and
\begin{equation}
U_{t+1} = \left ( 1 + \Gamma_L - {\Gamma_0 \over U_t} \right )^2 ,
 \hspace{1.0cm} U_0 = 1.
\label{eq:fracrecur-lin'}
\end{equation}

\noindent The fixed point equations are again cubic:
\begin{equation}
\left (\Gamma^* \right )^3 - 2 \left ( 1 + \Gamma_L \right )
 \left (\Gamma^* \right )^2
 + \left ( 1 + \Gamma_L \right )^2 \Gamma^* - \Gamma_0 = 0,
\label{eq:stressfix-lin'}
\end{equation}

\begin{equation}
\left (U^* \right )^3 - \left ( 1 + \Gamma_L \right )^2 \left (U^* \right )^2
 + 2 \left ( 1 + \Gamma_L \right ) \Gamma_0 U^* - \Gamma_0^2 = 0
\label{eq:fracfix-lin'}
\end{equation}

\noindent and they have the following solutions:

\begin{eqnarray}
\Gamma^*_1 & = & {2 \over3} \left ( 1 + \Gamma_L \right )
 + 2 K' \cos\frac{\Phi'}{3},
\label{eq:stressfix-lin1'} \\
\Gamma^*_2 & = & {2 \over 3} \left ( 1 + \Gamma_L \right )
 - K' \cos\frac{\Phi'}{3} + \sqrt{3} K' \sin\frac{\Phi'}{3},
\label{eq:stressfix-lin2'} \\
\Gamma^*_3 & = & {2 \over 3} \left ( 1 + \Gamma_L \right )
 - K' \cos\frac{\Phi'}{3} - \sqrt{3} K' \sin\frac{\Phi'}{3},
\label{eq:stressfix-lin3'}
\end{eqnarray}

\noindent where
\begin{equation}
K' = {1 + \Gamma_L \over 3},
\label{eq:K'-defn}
\end{equation}

\begin{equation}
\cos \Phi' = {27 \Gamma_0 \over 2 \left ( 1 + \Gamma_L \right )^3} - 1
\label{eq:Phi'-defn}
\end{equation}

\noindent and

\begin{eqnarray}
U^*_1 & = & {(1 + \Gamma_L)^2 \over 3} + 2 J' \cos\frac{\Theta'}{3},
\label{eq:fracfix-lin1'} \\
U^*_2 & = & {(1 + \Gamma_L)^2 \over 3} - J' \cos\frac{\Theta'}{3}
 + \sqrt{3} J' \sin\frac{\Theta'}{3},
\label{eq:fracfix-lin2'} \\
U^*_3 & = & {(1 + \Gamma_L)^2 \over 3} - J' \cos\frac{\Theta'}{3}
 - \sqrt{3} J' \sin\frac{\Theta'}{3},
\label{eq:fracfix-lin3'}
\end{eqnarray}

\noindent where
\begin{equation}
J' = \frac{1}{3} \sqrt{(1 + \Gamma_L)^4 - 6 (1 + \Gamma_L) \Gamma_0},
\label{eq:J'-defn}
\end{equation}

\begin{equation}
\cos\Theta' = {\left ( 1 + \Gamma_L \right )^3 \left [
 \left ( 1 + \Gamma_L \right )^3 - 9 \Gamma_0 \right ] + 27 \Gamma_0^2 / 2
 \over \left [ \left ( 1 + \Gamma_L \right )^4
               - 6 \left ( 1 + \Gamma_L \right ) \Gamma_0 \right ]^{3/2}}.
\label{eq:Theta'-defn}
\end{equation}

\begin{figure}[htb]
\resizebox{13.0cm}{!}{\rotatebox{0}{\includegraphics{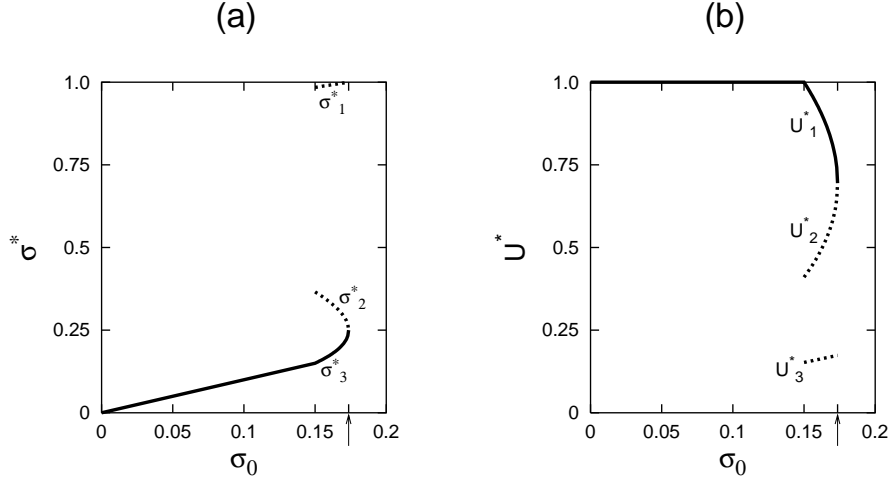}}}
\caption{The fixed points of (a) the redistributed stress,
and (b) the surviving fraction of fibers for the probability distribution
of fiber strengths shown in Fig.~\ref{fig6}. The curve for the stable
fixed points is shown by a bold solid line and those for the unstable
fixed points are shown by bold broken lines. In this example too
we have $\sigma_L = 0.15$ and $\sigma_R = 0.75$; here
$\sigma_0^{\rm crit} = 0.173611$, marked by an arrowhead. The
critical point is located lower than that in Fig.~\ref{fig5} due to
abundance of fibers of lower strengths compared to the previous case.}
\label{fig7}
\end{figure}

\noindent Here $\Gamma^*_3$ and $U^*_1$ are stable fixed points while
the rest are unstable (Fig.~\ref{fig7}).

\indent The discriminants of Eq.~(\ref{eq:stressfix-lin'}) and
Eq.~(\ref{eq:fracfix-lin'}) show that the critical applied stress in this
case, $\sigma_0^{\rm crit}\:'$ (or, $\Gamma_0^{\rm crit}\:'$),
is given by:

\begin{equation}
\Gamma_0^{\rm crit}\:'
 = {\sigma_0^{\rm crit}\:' \over \sigma_R - \sigma_L}
 = {4 \over 27} \left ( 1 + \Gamma_L \right ) ^3
\label{eq:stresscrit-lin'}
\end{equation}

\noindent or,
\begin{equation}
\sigma_0^{\rm crit}\:'
 = {4 \sigma_R^3 \over 27 \left ( \sigma_R - \sigma_L \right )^2}.
\label{eq:stresscrit-lin'-alter}
\end{equation}

\noindent In order to satisfy the condition $\sigma_0^{\rm crit}\:'
\geq \sigma_L$, it requires from Eq.~(\ref{eq:stresscrit-lin'-alter})
that

\begin{equation}
\sigma_R \geq 3 \sigma_L,
\label{eq:weak-strong-relate-lin'}
\end{equation}

\noindent which imposes an upper bound:

\begin{equation}
\sigma_0^{\rm crit}\:' \le {\sigma_R \over 3}.
\label{eq:stresscrit-lin'-up}
\end{equation}

\indent Like before, for $\Gamma_0 \leq \Gamma_0^{\rm crit}\:'$
the stable fixed points are real-valued, which indicates that only partial
failure of the fiber bundle takes place before a state of mechanical
equilibrium is reached; for $\Gamma_0 > \Gamma_0^{\rm crit}\:'$
the fixed points are not real and a phase of total failure exists.
The order parameter $\cal O$ of the transition is given by the
definition in Eq.~(\ref{eq:order-lin-defn}).

\noindent For $\Gamma_0 = \Gamma_0^{\rm crit}\:'$ we get the
following properties from Eq.~(\ref{eq:Phi'-defn}),
Eq.~(\ref{eq:fracfix-lin1'}), Eq.~(\ref{eq:fracfix-lin2'}) and
Eq.~(\ref{eq:Theta'-defn}):

\begin{equation}
U^*_{1-{\rm crit}} = U^*_{2-{\rm crit}}
 = {4 \over 9} \left ( 1 + \Gamma_L \right )^2;
\label{eq:fracfix-lin12'-crit}
\end{equation}

\begin{equation}
\cos \Theta'_{\rm crit} = -1 \hspace{1.0cm}
 {\rm or}, \hspace{1.0cm} \Theta'_{\rm crit} = \pi
\label{eq:Theta'-crit}
\end{equation}

\noindent and
\begin{equation}
\cos \Phi'_{\rm crit} = 1 \hspace{1.0cm}
 {\rm or}, \hspace{1.0cm} \Phi'_{\rm crit} = 0.
\label{eq:Phi'-crit}
\end{equation}

\noindent Comparing Eq.~(\ref{eq:Phi'-crit}) and Eq.~(\ref{eq:Theta'-crit})
with Eq.~(\ref{eq:Phi-Theta-crit2}) we see that the critical values of
$\Theta$ and $\Theta'$ are the same whereas those of $\Phi$ and
$\Phi'$ differ by $\pi$ radians.

\indent Near the critical point, but below it, we get from
Eq.~(\ref{eq:fracfix-lin1'}) and Eq.~(\ref{eq:fracfix-lin2'}):

\begin{equation}
U^*_1 \simeq U^*_{\rm 1-crit}
 + {4 \over 3} \left (1 + \Gamma_L \right ) ^{1/2}
 \left ( \Gamma_0^{\rm crit}\:' - \Gamma_0 \right )^{1/2}
\label{eq:fracfix-lin1'-nearcrit}
\end{equation}

\noindent and
\begin{equation}
U^*_2 \simeq U^*_{\rm 2-crit}
 - {4 \over 3} \left (1 + \Gamma_L \right ) ^{1/2}
 \left ( \Gamma_0^{\rm crit}\:' - \Gamma_0 \right )^{1/2}.
\label{eq:fracfix-lin2'-nearcrit}
\end{equation}

\noindent Therefore, by the definition of the order parameter in
Eq.~(\ref{eq:order-lin-defn}) and that of the susceptibility in
(\ref{eq:suscept-lin-crit}), we get in this case ${\cal O} \propto\left (
\Gamma_0^{\rm crit}\:' - \Gamma_0 \right )^{1/2}$ and
$\chi \propto \left ( \Gamma_0^{\rm crit}\:' - \Gamma_0 \right )^{-1/2}$,
$\Gamma_0 \to \Gamma_0^{\rm crit} -$. These power laws have the
same exponents as the corresponding ones in the previous cases and
differ from those only in the critical point and the critical amplitude.

\indent At the critical point the asymptotic relaxation of the
surviving fraction of fibers to its stable fixed point [obtained as
an asymptotic solution to (\ref{eq:fracrecur-lin'})] is again found to be
a power law decay similar to Eq.~(\ref{eq:frac-uni-crit-dyn})
and Eq.~(\ref{eq:frac-lin-crit-dyn}):

\begin{eqnarray}
U_t - U^*_{\rm 1-crit} & \sim &
 {4 \over 3} {U^*_{\rm 1-crit} \over t}, \hspace{1.0cm} t \to \infty
 \nonumber \\
 & \sim & {16 \over 27} \left ( 1 + \Gamma_L \right )^2 {1 \over t}.
\label{eq:frac-lin-crit-dyn'}
\end{eqnarray}

\indent The critical behaviour of the models reported in this and the
previous two sections show that the power laws found here
are independent of the form of the probability distribution $P$.
The three probability distributions studied have a common
feature: the function $\sigma^* \left [ 1 - P \left( \sigma^* \right)
\right ]$ has a maximum in the interval $\left ( \sigma_L, \sigma_R
\right )$ which corresponds to the critical value of the initial
applied stress. All probability distributions having this property
are therefore expected to lead to the same universality class as
the three studied here. If the probability distribution does not
have this property we may not observe a phase transition at all.
For example, consider a fiber bundle model with
$P(\sigma_{\rm thresh}) = 1 - 1/\sigma_{\rm thresh}$,
$\sigma_{\rm thresh} \geq 1$.
Here $\sigma^* \left [ 1 - P \left( \sigma^* \right) \right ] = 1$
and the evolution of the fiber bundle is given by the recursion
relation $U_{t+1} = U_t/\sigma_0$ which implies that there is
no dynamics at all for $\sigma_0 = 1$ and an exponential decay
to complete failure, $U_t = (\sigma_0)^{-t}$, for $\sigma_0 > 1$.
There are no critical phenomena and therefore no phase transition.
However this general conclusion may not be true for finite-sized
bundles~\cite{McCartney1983}.

\section{Discussion}

\indent In this paper we have studied the critical properties of
failure in a class of fiber bundle models under an applied stress.
The models are simple dynamical systems that show an irreversible
phase transition. We have determined the static and dynamic critical
properties associated with the phase transition. Since the models
have been defined without any fluctuations in the local density
of fiber strengths or in the load sharing, these are equivalent to a
mean-field theory. We have defined a new order parameter in
Eq.~(\ref{eq:order-uni-defn}) which shows that the transition is of
second-order. It is supported by facts which are characteristic of
second-order transitions: the susceptibility diverges at the critical
point and the decay of surviving fraction of fibers with time at
the critical point follows a power-law. Besides, the ralaxation time
of the bundle diverges from both sides of the critical point:
for an initial stress infinitesimally below the critical point
the surviving fraction of fibers will take an infinite time
to reach its stable fixed point, whereas infinitesimally above the
critical point it will take an infinite time to get past the
fixed point.
By obtaining the same critical exponents for three different
probability distrbutions of fiber strengths, we are inclined to
conclude that these exponents are universal. The universality
of the critical exponent for susceptibility was also reported in
Ref.~\cite{daSilveira1998, daSilveira1999}.

\indent The two models studied in sections 4 and 5 seem to be
related by the fact that their density functions,
Eq.~(\ref{eq:dens-lin}) and Eq.~(\ref{eq:dens-lin'}),
can be transformed from one to the other by a reflection on the
line $\sigma_{\rm thresh} = (\sigma_L + \sigma_R)/2$
(compare Fig.~\ref{fig4}(a) and Fig.~\ref{fig6}(a)).
But the fixed point equations and their
solutions do not have this symmetry. This is because
the density function $\rho(\sigma_{\rm thresh})$ does not appear
directly in the recursion relations for the dynamics. It is the
distribution function $P(\sigma_{\rm thresh})$ which appears in
the recursion relations. Eq.~(\ref{eq:prob-lin}) and
Eq.~(\ref{eq:prob-lin'}) show that the distribution functions of
these two models are not mutually symmetric about any value of
the threshold stress $\sigma_{\rm thresh}$
(compare Fig.~\ref{fig4}(b) and Fig.~\ref{fig6}(b)).
However a certain relation exists between the critical
values of the applied stress for a special case of these two models:
if $\sigma_L = 0$, we get from Eq.~(\ref{eq:stresscrit-lin}) and
Eq.~(\ref{eq:stresscrit-lin'}) that $\sigma_0^{\rm crit} / \sigma_R
 = \sqrt{4/27}$ and $\sigma_0^{\rm crit}\:' / \sigma_R = 4/27$
respectively; therefore we have $\sigma_0^{\rm crit}\:' / \sigma_R
 = \left ( \sigma_0^{\rm crit} / \sigma_R \right )^2$.

\indent Finally we compare the fiber bundle model studied in
this paper with the mean-field Ising model. Though the order parameter
exponent (equal to $\frac{1}{2}$) of this model is identical to
that of the mean-field Ising model the two models are not in the
same universality class. The susceptibility in these models diverge
with critical exponents $\frac{1}{2}$ and $1$ respectively on
approaching the critical point. The dynamical critical exponents
are not the same either: in this fiber bundle model the surviving
fraction of fibers under the critical applied stress decays toward
its stable fixed point as $t^{-1}$, whereas the magnetization of the
mean-field Ising model at the critical temperature decays to zero
as $t^{-1/2}$.

\end{document}